\documentclass[10pt,journal,compsoc,twocolumn]{IEEEtran}

\usepackage{booktabs}
\usepackage{multirow}
\usepackage{flushend}
\usepackage{subfigure}
\usepackage{hyperref}
\usepackage{dsfont,amsfonts,amssymb,amsmath,color}
\usepackage{latexsym}
\usepackage{amsmath}
\usepackage{amsfonts}
\usepackage{algorithm}
\usepackage[noend]{algpseudocode}
\usepackage{graphicx}
\usepackage{color}
\usepackage{algorithm}  
\usepackage{algpseudocode}
\usepackage{amsmath}  
\begin{document}

\title{Hierarchical Taxonomy-Aware and Attentional Graph Capsule RCNNs for Large-Scale Multi-Label Text Classification}

\author{Hao Peng,
        Jianxin Li\emph{, Member, IEEE},
        Qiran Gong,
        Senzhang Wang,
        Lifang He,\\
        Bo Li,
        Lihong Wang, and Philip S. Yu\emph{, Fellow, IEEE},
\IEEEcompsocitemizethanks{
\IEEEcompsocthanksitem Hao Peng, Jianxin Li, Qiran Gong and Bo Li are with  Beijing Advanced Innovation Center for Big Data and Brain Computing, Beihang University, Beijing 100083, and also with the State Key Laboratory of Software Development Environment, Beihang University, Beijing 100083, China, China. E-mail:\{penghao, lijx, libo\}@act.buaa.edu.cn, allen\_gong@buaa.edu.cn.
\IEEEcompsocthanksitem Senzhang Wang is with the College of Computer Science and Technology, Nanjing University of Aeronautics and Astronautics, Nanjing 211106, China. E-mail: szwang@nuaa.edu.cn.
\IEEEcompsocthanksitem Lifang He is with the Department of Biostatistics and Epidemiology, University of Pennsylvania, Philadelphia, PA 19104, USA. E-mail: lfhe@pennmedicine.upenn.edu.
\IEEEcompsocthanksitem Lihong Wang is with the National Computer Network Emergency Response Technical Team/Coordination Center of China, Beijing 100029, China. E-mail: wlh@isc.org.cn.
\IEEEcompsocthanksitem Philip S. Yu is with the Department of Computer Science, University of Illinois at Chicago, Chicago, IL 60607, USA. E-mail: psyu@uic.edu.
}
\thanks{ Manuscript received May 6, 2019. (Corresponding author: Jianxin Li.)}
}

\IEEEtitleabstractindextext{%
\begin{abstract}
CNNs, RNNs, GCNs, and CapsNets have shown significant insights in representation learning and are widely used in various text mining tasks such as large-scale multi-label text classification.
However, most existing deep models for multi-label text classification consider either the non-consecutive and long-distance semantics or the sequential semantics, but how to consider them both coherently is less studied.
In addition, most existing methods treat output labels as independent medoids, but ignore the hierarchical relations among them, leading to useful semantic information loss.
In this paper, we propose a novel hierarchical taxonomy-aware and attentional graph capsule recurrent CNNs framework for large-scale multi-label text classification. 
Specifically, we first propose to model each document as a word order preserved graph-of-words and normalize it as a corresponding words-matrix representation which preserves both the non-consecutive, long-distance and local sequential semantics.
Then the words-matrix is input to the proposed attentional graph capsule recurrent CNNs for more effectively learning the semantic features.
To leverage the hierarchical relations among the class labels, we propose a hierarchical taxonomy embedding method to learn their representations, and define a novel weighted margin loss by incorporating the label representation similarity.
Extensive evaluations on three datasets show that our model significantly improves the performance of large-scale multi-label text classification by comparing with state-of-the-art approaches.
\end{abstract}

}

\maketitle

\section{Introduction}
As a fundamental text mining task, text classification aims to assign a text with one or several category labels such as topic labels and sentiment labels.
Traditional approaches represent the text as sparse lexical features due to the simplicity and effectiveness ~\cite{Aggarwal2012}.
For example, bag-of-words and n-gram are widely used to extract textual features, and then a general machine learning model such as Bayesian, logistic regression or SVM is utilized for text classification.
With the development of deep learning techniques ~\cite{Lecun2015Deep,goodfellow2016deep}, variants of neural network based models have been exploited from a large body of innovations, such as recurrent neural networks~\cite{Tai2015Improved,Wang2016Attention,Yang2017Hierarchical,shen2018biblosan}, diversified convolutional neural networks~\cite{Kim2014Convolutional,YaoGCN2018,Liu:2017:DLE:3077136.3080834,Peng:2018,Conneau2016Very}, capsule neural networks~\cite{Zhao2018Investigating} and adversarial structures ~\cite{Liu2017Adversarial,miyato2016adversarial}. 
These deep models have achieved inspiring performance gains on text classification due to their powerful capacity in representing the text as a fix-size feature map with rich semantics information.

Recently, three popular deep learning architectures have attracted increasing research attention for text data, i.e., recurrent neural networks (RNNs)~\cite{Yang2017Hierarchical,TangQL15,shen2018biblosan,shen2018disan}, convolutional neural networks (CNNs)~\cite{Kim2014Convolutional,Conneau2016Very,Liu:2017:DLE:3077136.3080834} and graph convolutional networks (GCNs)~\cite{Peng:2018,YaoGCN2018}.
RNNs are more powerful on capturing the semantics of short text~\cite{bengio2003neural}, but are less effective to learn semantic features of long documents.
Although the bi-directional block self-attention networks are proposed~\cite{shen2018biblosan} to better model text or sentence, they consider documents as natural sequences of words, and ignore the long-distance semantic between paragraphs or sentences.
CNNs simply evaluate the semantic composition of the consecutive words extracted with n-gram, while n-gram may lose the long-distance semantic dependency among the words~\cite{Aggarwal2012}.
Compared with RNNs and CNNs, GCNs can better capture the non-consecutive phrases and long-distance word dependency semantics~\cite{Peng:2018,YaoGCN2018}, but ignore the sequential information.
To sum up, there still lacks of a model that can simultaneously capture the non-consecutive, long-distance and sequential semantics of text. 
Meanwhile, as the text labels of some real-world text classification tasks are characterized by large hierarchies, there may exist strong dependency among the class labels~\cite{sun2001hierarchical,xue2008deep,Gopal2012Bayesian}.
Existing deep learning models cannot effectively and efficiently leverage the hierarchical dependencies among labels for improving the classification performance, either.

It is non-trivial to obtain a desirable classification performance for large-scale multi-label text due to the following major challenges.
First, although there are many methods for document modeling, how to represent a document by fully preserving its rich and complex semantic information still remains an open problem~\cite{Berry:2003:STM:945832}.
It is challenging to come up with a document modeling method that can fully capture the semantics of a document, including the non-consecutive, long-distance and sequential semantics of the words.
Second, existing CNNs, RNNs and GCNs models usually can only capture partial textual features. 
It is challenging to design a deep learning model that can simultaneously capture multiple types of textual features mentioned above.
Third, although some recursive regularization based hierarchical text classification models ~\cite{Gopal:2015:HBI:2737800.2629585,Gopal2013Recursive,Peng:2018,xie2013multilabel} consider the pair-wise relation between labels, they fail to consider their hierarchical relations.
In addition, the computation of the above regularized models is expensive due to the use of Euclidean constraints.
How to make full use of the hierarchical label-dependencies among labels to improve the classification accuracy and reduce the computational complexity is also challenging.

To address the above challenges, we propose a novel \underline{H}ierarchical taxonomy-awar\underline{E} and \underline{A}ttentional \underline{G}raph \underline{C}apsule \underline{R}ecurrent \underline{CNN}s framework called HE-AGCRCNN for large-scale multi-label text classification. 
Specifically, our framework contains three major parts: word order preserved graph-of-words for document modeling, attentional capsule recurrent CNNs for features learning, and hierarchical taxonomy-aware weighted margin loss for multi-label text classification.
Next we will elaborate the three parts as follows.

\textbf{Word Order Preserved Graph-of-Words for Document Modeling.}
We regard each word as a vertex, the word co-occurrence relationships within a sliding window as edges, and the positional index of a word appearing in the document as its attribute.
In this way, we build a word order preserved graph-of-words to represent a document.
Then we select top $N$ central words from the graph-of-words based on the \emph{closeness centrality}, and construct a subgraph for each central word from neighbors by breadth first search (BFS) and depth first search (DFS).
To preserve local sequential, non-consecutive and long-distance semantics, we next normalize each subgraph to blocks of word sequences that retains local word order information by utilizing the attribute of the vertex, and construct an arranged words-matrix for the $N$ sub-graphs.
To incorporate more semantic information, we use a pre-trained word embedding vectors based on word2vec ~\cite{Mikolov:2013:DRW:2999792.2999959,Mikolov2013Efficient} as word representation in the arranged words-matrix. 
Finally, each document is represented as a corresponding 3-D tensor whose three dimensions are the selected central words, the ordered neighbor words sequence, and the embedding vector of each word, respectively.

\textbf{Attentional Capsule Recurrent Convolutional Neural Networks.}
An attentional capsule recurrent CNN (RCNN) model is designed to make use of the document tensor as input for document features learning. 
The proposal model first uses two attentional RCNN layers to learn different levels of text features with both non-consecutive, long-distance and local sequential semantics.
Here, we not only guarantee the independence of the feature representation between sub-graphs, but also model different impacts among different blocks of word sequences.
When the convolution kernel slides horizontally along the combining long-distance and local sequential ordering of words, the attentional RNN unit is used to encode the output of the previous step of CNN, and the output of current step of attentional RNN to produce the final output feature map in the RCNNs layer.
Then a capsule network layer is used to implement an iterative routing process to learn the intrinsic spatial relationship between text features from lower to higher levels for each sub-graph.
In the final DigitCaps layer, the activity vector of each capsule indicates the presence of an instance of each class and is used to calculate the classification loss.

\textbf{Hierarchical Taxonomy-Aware Weighted Margin Loss.}
Considering the hierarchical taxonomy of the labels, we design two types of meta-paths, and use them to conduct random walk on the hierarchical label taxonomy network to generate label sequences.
Therefore, the hierarchical taxonomy relation among the labels can be encoded in a continuous vector space with the skip-gram~\cite{Mikolov2013Efficient} on the sequences.
In this way, the distance between two labels can be measured by calculating the cosine similarity of their label vectors. 
By taking the distance between labels into consideration, we design a new weighted margin loss to guide the training of proposed models in multi-label text classification.

We conduct extensive evaluations on our proposed framework by comparing it with state-of-the-art methods on three benchmark datasets, comparing with traditional shallow models and recent deep learning models. 
The results show that our approach outperforms them by a large margin in both efficiency and effectiveness on large-scale multi-label text classification.

The contributions of this paper are summarized below.
\begin{itemize}
\item A novel hierarchical taxonomy-aware and attentional graph capsule recurrent CNNs framework is proposed for large-scale multi-label text classification.
\item A new word order preserved graph-of-words method is proposed to better model document and more effectively extract textual features. The new document modeling method preserves both non-consecutive, long-distance and local sequential semantics.
\item A new word sequence block level attention recurrent neural network is proposed to better learn local sequential semantics of text.
\item A novel hierarchical taxonomy-aware weighted margin loss is proposed to better measure the distance of classes in hierarchy and guide the proposed models training.
\item Extensive evaluations on three benchmark datasets demonstrate the efficiency and effectiveness of the proposal.
\end{itemize}

\begin{figure*}[h]
\center
\includegraphics[width=0.9\textwidth]{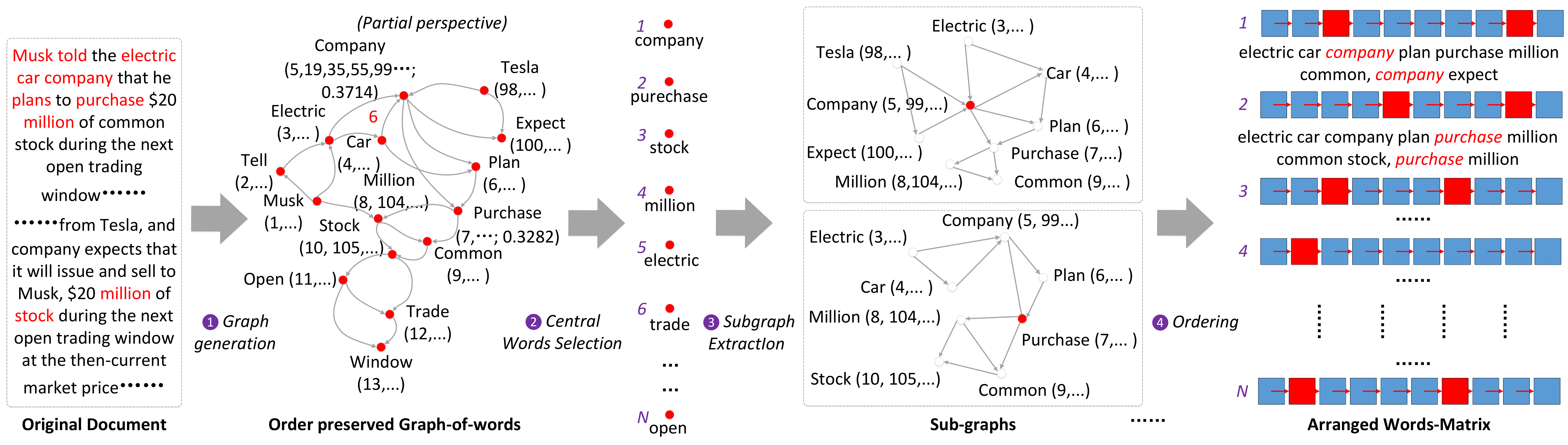}\vspace{-0.15in}
\caption{Illustration of converting a document to an arranged words-matrix representation. We first construct a word order preserved graph-of-words from the original document, and then a top $N$ nodes (words) sequence is selected from the ranking of each node's \emph{closeness centrality} feature. For each node (word) in the sequence, a corresponding sub-graph is extracted and normalized as a sequence of words that retain local word order information.}\label{fig:text2norma}\vspace{-0.2in}
\end{figure*}

The rest of the paper is organized as follows.
We first introduce the word order preserved graph-of-words based document modeling in Section~\ref{sec:graphfordoc}.
Then we present the model architecture in Sections~\ref{sec:dgcrcnn} and ~\ref{sec:labelembedding}. 
The evaluation is conducted in Section~\ref{sec:experi}. 
Finally, we review related work in Section~\ref{sec:relatedwork} followed by the conclusion and future work in Section~\ref{sec:conclu}.

\section{Word order Preserved graph-of-words for Document Modeling}\label{sec:graphfordoc}
In this section, we introduce how we model a document as a word order preserved graph-of-words, and how to extract central words and sub-graphs from it to preserve both non-consecutive, long-distance and local sequential semantics of the document.
Formally, we denote the training document set as $\mathcal{D} = \{d_{s} ,T_{s}\}_{s=1}^{M}$, where $M$ is the total number of documents in $\mathcal{D}$, $d_{s}$ is a document, $T_{s}$ is the label set of $d_{s}$ and $T_{s} \subset \mathcal{S}$. 
We also denote the set of labels as $\mathcal{S} = \{v_{i}|i=1, 2, \cdots, L\}$, where $L$ is the total number of labels.

\subsection{Word Order Preserved Graph-of-Words Construction}
In order to preserve more semantic information of text, we model a document as a word order preserved graph-of-words.
We regard each word as a vertex, the word co-occurrence relationships within a sliding window as edges, and the positional index appearing in the document as its attribute, as shown in the step 1 of Figure~\ref{fig:text2norma}.

We first split a document into a set of sentences and extract tokens using Stanford CoreNLP tool\footnote{http://stanfordnlp.github.io/CoreNLP/}.
We also employ a lemmatization of each token using Stanford CoreNLP, and remove the stop words. 
Then we construct an edge between two word nodes if they co-occur in a pre-defined fixed-size sliding window, and the weight of the edge is the times of their co-occurrence. 
Meanwhile, we record all the positional indexes where a word appears in the document as its attribute.
For example, for the first sentence ``\emph{Musk told the electric car company that...}'' shown in the document of Figure ~\ref{fig:text2norma}, we perform lemmatization on the second word ``\emph{told}'' to get ``\emph{tell}'' with attribute ``2'', and build a directed edge from ``\emph{Musk}'' to each of the words in the sliding window.
As shown in the word order preserved graph-of-words of Figure ~\ref{fig:text2norma}, the word ``\emph{Company}'' appears at the $5$-th, $19$-th, $35$-th, $55$-th, $99$-th, etc. positions, respectively.
Note that the word order preserved graph-of-words is a weighted directed graph with the positional indexes as the node attributes.
For example, in the word order preserved graph-of-words of Figure ~\ref{fig:text2norma}, the weight of the edge between nodes ``\emph{Company}'' and  ``\emph{Car}'' is $6$ meaning that ``\emph{Company}'' and  ``\emph{Car}'' has a total of 6 co-occurrences in the sliding window.

\subsection{Arranged Words-Matrix Generation}
We denote the word order preserved graph-of-words as $\mathcal{G} = (V, E, W, A)$, where $V$ denotes the node set and $|V| = n$, $E$ denotes the edge set and $|E| = m$, $W$ denotes the weights of the edges and $A$ denotes the attributes of the nodes. 
We extract top $N$ central words from $\mathcal{G}$ based on node's \emph{closeness centrality} feature.
Here, in order to calculate \emph{closeness centrality} for each node, we use $d(v, u)$ to denote the shortest-path distance between nodes $v$ and $u$ by using the Dijkstra algorithm.
For each node $v$, its \emph{closeness centrality} can be calculated by $C_{v} ={(n-1)}/{\sum_{u\in{V}, u\neq v}d(v,u)}$.
So we can arrange the nodes in order of largest to smallest according to their \emph{closeness centrality} features.
The larger the \emph{closeness centrality}, more important the node is in the graph.
As shown in the word order preserved graph-of-words of Figure ~\ref{fig:text2norma}, the \emph{closeness centrality} of word ``\emph{Company}'' is the highest $0.3714$ among the words' in the graph.
Then we select the top $N$ central nodes from the node ``\emph{Company}'' to the node ``\emph{Open}'', as shown in the step 2 of Figure~\ref{fig:text2norma}. 
Next, we introduce how to extract sub-graph $\mathcal{G}(v)$ for each selected central node $v$.

\begin{figure*}[t]
\center
\includegraphics[width=0.9\textwidth]{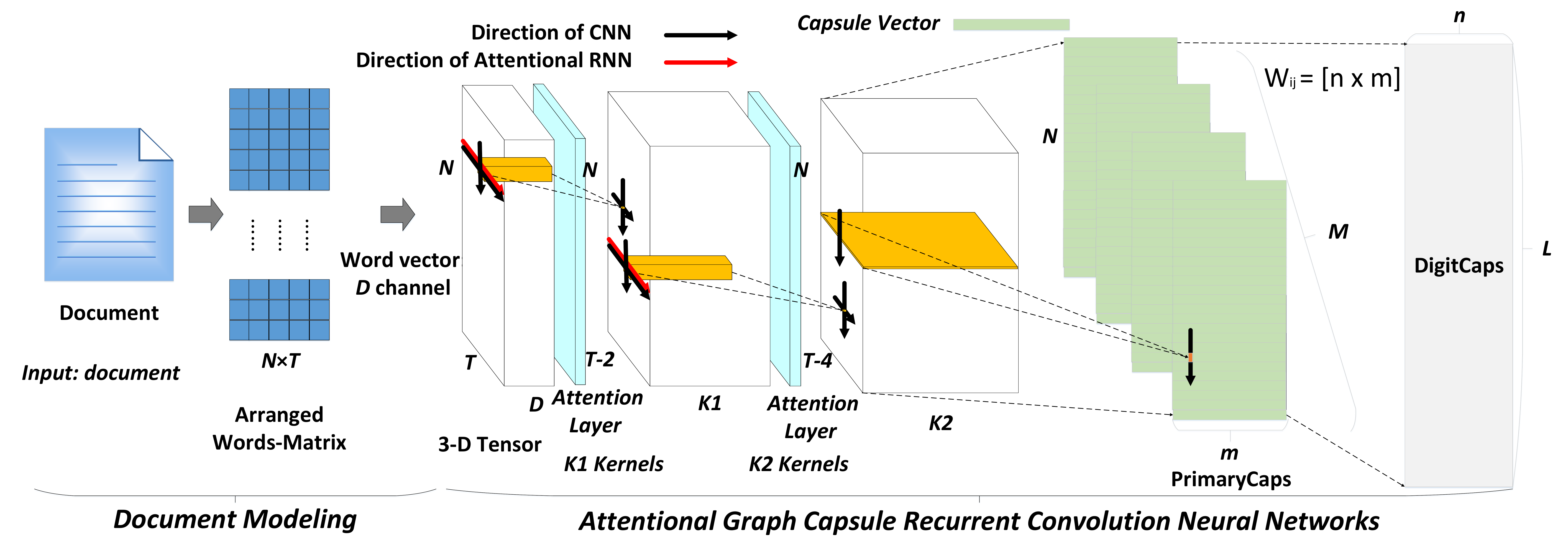}\vspace{-0.15in}
\caption{Architecture of the proposed hierarchical taxonomy-aware and attentional graph capsule recurrent convolution neural network. It consists of document modeling, attentional capsule recurrent CNN, and hierarchical taxonomy-aware weighted margin loss for multi-label text classification. The network input is the original document. The length of the activity vector of each capsule in DigitCaps layer indicates presence of an instance of each class.
}\label{fig:deepcrcnn}\vspace{-0.2in}
\end{figure*}

First, we extract the nodes and edges from the neighborhood of each central node in the order of breadth first search (BFS), depth first search (DFS) and the node's \emph{closeness centrality} feature to build a subgraph.
Meanwhile, we limit the number of nodes in the subgraph to be no more than $K$, as shown in the step 3 of Figure~\ref{fig:text2norma}.
In this way, the sub-graph $\mathcal{G}(v)$ contains both the non-consecutive, long-distance and local sequential information of the central word $v$ in the document. 
To further save the above information of a subgraph, we order the words in the sub-graph $\mathcal{G}(v)$ by their nodes (words) attributes.
For the most case where a subgraph contains multiple sentences, we guarantee that the long sentences are in the front and the short sentences are in the back.
As shown in the first line of the arranged words-matrix in Figure~\ref{fig:text2norma}, we convert the first sub-graph $\mathcal{G}(v)$ as sequences liking``\emph{electric car company plan purchase million common} and \emph{company expect }''.
As a result, we normalize each subgraph as sequences of nodes (words) that keep the same length $T$.
If the number of words in the sequence is less than $T$, it is padded with zeros.
Finally, we concatenate all the normalized sequences of the $N$ central words into an arranged words-matrix, as shown in step 4 of Figure~\ref{fig:text2norma}.
The red nodes represent central words.


\subsection{Unified Representation of the Documents}
For better representing the original words in the words-matrix, we use word2vec~\cite{Mikolov2013Efficient,Mikolov:2013:DRW:2999792.2999959} to incorporate as much word semantic information as possible. 
Specifically, word2vec is trained on a larger corpus, i.e., Wikipedia.
All parameters for word2vec are set to be default values. 
In this way, we have a 3-D tensor representation for each document, where the padded vectors are zero vectors with the same dimension. 
Then the convolution, recurrent and capsule networks introduced in the next section will be operated over the unified representations of the documents.

\section{Attentional Capsule Recurrent CNN}\label{sec:dgcrcnn}
In this section, we introduce the proposed attentional capsule recurrent CNN model. 
After converting each document into a 3-D tensor representation, we design a three layers of attentional capsule recurrent CNN model to learn both the non-consecutive, long-distance and local sequential feature.
From the input document to the output labels, the architecture is shown in Figure~\ref{fig:deepcrcnn}.
Specifically, the three layers of neural networks contain two major parts: two layers of attentional recurrent convolution neural networks and one layer of capsule networks for rich feature learning.
Note that this is a general framework and the number of attentional recurrent convolution layers can be adjusted based on specific dataset for classification, and the parameter configuration of self-attentional recurrent operators and capsule networks can be customized in different text classification tasks.

\subsection{Attentional Recurrent CNN}
Different from the architecture of existing recurrent convolutional neural networks~\cite{Lai:2015:RCN:2886521.2886636}, which encode sentences or document as a dense vector for classification, our proposed attentional recurrent convolution neural networks encode whole document as 3-D feature map.
The attentional recurrent CNN model takes the $N\times T \times D$ size of 3-D tensor extracted from the document and word embedding as the input, where $N$ is the number of central words, $T$ is the length of normalized sequence of words and $D$ is the dimension of word embedding, as shown in Figure~\ref{fig:deepcrcnn}.
The output of the two layers of attentional recurrent convolution networks is the other 3-D feature map as input of the proposed capsule network.

\begin{figure}[t]
\center
\includegraphics[width=0.5\textwidth]{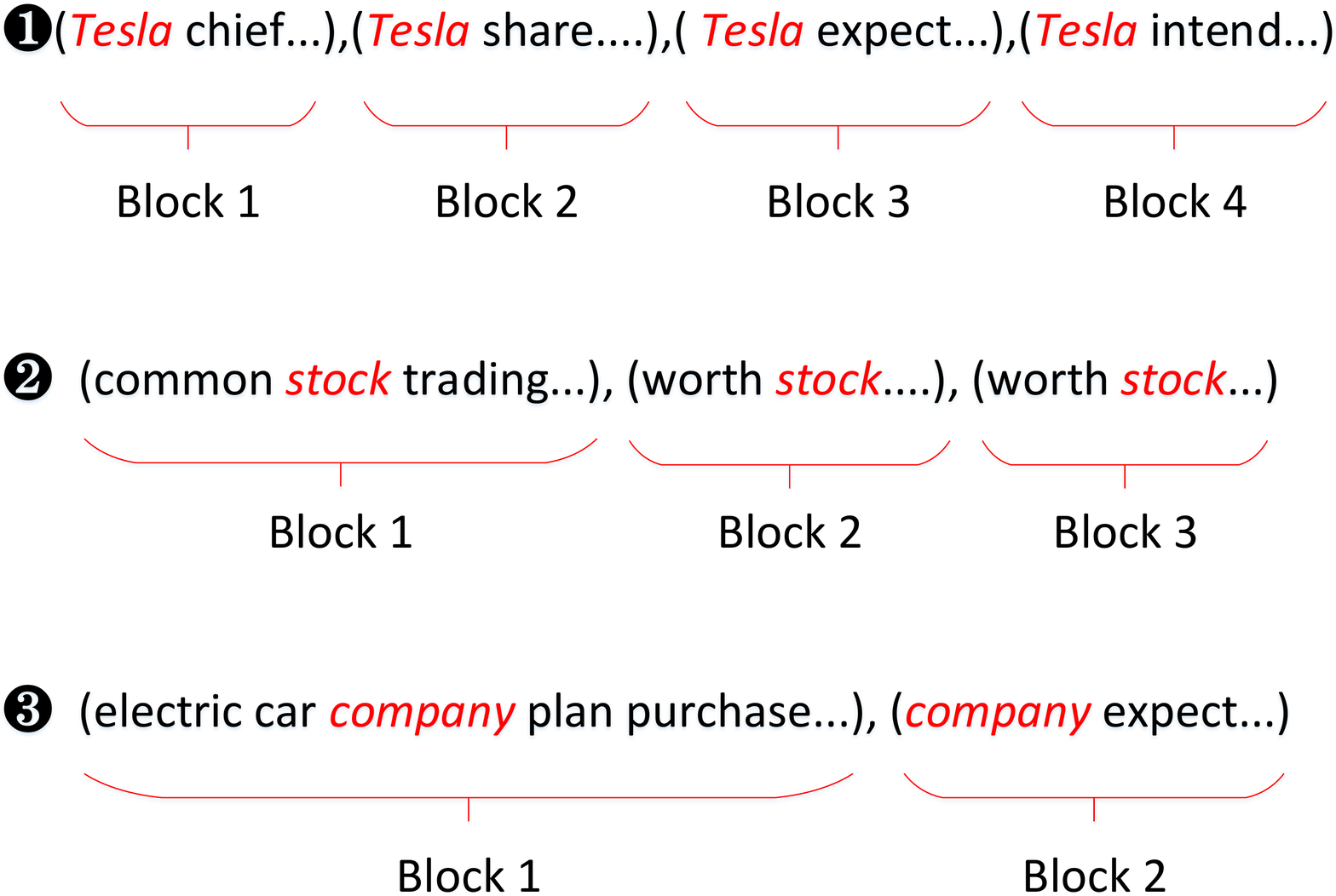}\vspace{-0.2in}
\caption{Illustration of blocks of word sequences. The red word refers to the central word for each subgraph. The words in any block are the contexts of the central word at different locations in a document.}\label{fig:blocks}\vspace{-0.2in}
\end{figure}

In the first layer, the convolution operator filters the input tensor with $k1$ kernels of size $1\times 3 \times D$ with a horizontal stride of 2 elements and a vertical stride of 1 element, which is illustrated with the black convolution slide direction arrow in Figure~\ref{fig:deepcrcnn}. 
We use \emph{ReLU} as the activation function to speed up the training process and avoid over-fitting. 
Here, convolution kernel serves as a composition of the semantics in the receptive field to extract the higher level semantic features.
Meanwhile, we employ a masked attentional recurrent neural operator to capture the local sequential semantic for each sub-graph $\mathcal{G}(v)$.
The attentional recurrent neural operator acts on each horizontal words sequence, which is illustrated with the red recurrent slide direction arrow in Figure~\ref{fig:deepcrcnn}.
However, as we know, we convert the subgraph $\mathcal{G}(v)$ into blocks of word sequences according to the properties of the nodes.
We give three different blocks, as shown in Figure~\ref{fig:blocks}, to illustrate the blocks of word sequences, which consist of each line of the arranged words-matrix.
In order to measure the different impacts of different number of blocks on the local sequential semantic learning, we customize the masked attentional parameter shared long short-term memory, namely Attention-LSTM, to learn the rich local sequential semantic for each sub-graph $\mathcal{G}(v)$.
Since our proposed framework needs to learn feature for multiple documents, the attention-LSTM module guarantees that any subgraphs with the same order and number of blocks share the same attention parameters.
For example, for the $T$-th subgraph $\mathcal{G}(u)$ from the document $d_i$, assuming that it contains the $q$ blocks of word sequences, the parameter of the masked attention module is the $\alpha_{B_{T, q, 1}},\alpha_{B_{T, q, 2}}, \dots, \alpha_{B_{T, q, q}}$.
However, for the $T$-th subgraph $\mathcal{G}(v)$ from the document $d_j$, assuming that it contains the $p$ blocks of word sequences, the parameter of the masked attention module is the $\alpha_{B_{T, p, 1}},\alpha_{B_{T, p, 2}}, \dots, \alpha_{B_{T, p, p}}$.
But, among any one block, each word shares the same attention parameter.
For example, in Figure~\ref{fig:blocks}, we assume they are converted from the top 3 subgraphs.
In the $1$-th block of the $3$-th subgraph, the words \emph{electric}, \emph{car}, \emph{company}, \emph{plan}, and \emph{purchase} share the same masked attention parameter $\alpha_{B_{3,2,1}}$.
Then, after the first attentional recurrent convolution layer, there is an $N\times (T-2) \times k1$ size of feature map.
Compared with traditional convolution and recurrent networks on text data  ~\cite{Kim2014Convolutional,Conneau2016Very,Liu:2017:DLE:3077136.3080834,Peng:2018,Lai:2015:RCN:2886521.2886636,shen2018biblosan}, the significant difference of our designed attentional recurrent CNN units is that it can integrate the long-distance, non-consecutive and local sequential semantics of the corresponding sub-graph $\mathcal{G}(v)$.

The second attentional recurrent convolution layer takes the output of the first attentional recurrent convolution as its input, and filters it with $k2$ kernels of size $1\times 3 \times k1$ with a horizontal stride of 2 elements and a vertical stride of 1 element, which are illustrated with the black convolution slide direction arrow and red recurrent slide direction arrow in Figure~\ref{fig:deepcrcnn}.
We still guarantee that each horizontal feature map characterizes the semantics of corresponding sub-graph $\mathcal{G}(v)$, and the attentional recurrent and convolution operators between different sub-graphs are independent.
In the second layer, for each sub-graph $\mathcal{G}(v)$, the number of attentional parameter is same with the first layer, but they are separated in training. 
After the second attentional recurrent convolution layer's operation, a $N\times (T-4)\times k2$ size of feature map is generated.

More formally, we we give the definitions of convolution operator and Attentional-LSTM unit, respectively.
The convolution operator can be defined as
\begin{equation}\label{eq:cnn_operators}
\begin{aligned}
\centering
x_{j}^{l} = f(\sum_{i\in M_{j}}x_{i}^{l-1}\cdot k_{ij}^{l} + b_{j}^{l}),
\end{aligned}
\end{equation}
where $x_{j}^{l}$ represents the $j$-th feature map of the $l$-th layer of the convolution network, and $l\in\{1,2\}$.
This formula shows the convolution operation and the summation for all the associated feature maps $x_{i}^{l-l}$ and the $j$-th convolution kernel $k_{ij}^{l}$ of layer $l$, and then add an offset parameter $b_{j}^{l}$. 
Finally, a \emph{ReLU} activation function $f$ is applied.
Meanwhile, the Attentional-LSTM unit can be defined as:
\begin{equation}\label{eq:rnn_operators}
\begin{aligned}
\centering
f_{t} =& \sigma_{g}(W_{f}\alpha_{B}x_{t} + U_{f}c_{t-1} + b_{f}),\\
i_{t} =& \sigma_{g}(W_{i}\alpha_{B}x_{t} + U_{i}c_{t-1} + b_{i}),\\
o_{t} =& \sigma_{g}(W_{o}\alpha_{B}x_{t} + U_{o}c_{t-1} + b_{o}),\\
c_{t} =& f_{t}c_{t-1} + i_{t}\sigma_{c}(W_{c}\alpha_{B}x_{t} + b_{c}),\\
h_{t} =& o_{t}\sigma_{h}(c_{t}),
\end{aligned}
\end{equation}
where $t$ refers to the index of the horizontal convolution sequence, $f_{t}$ refers to the forgotten gate, $i_{t}$ refers to the input gate, $o_{t}$ refers to the output gate, $c_{t}$ is the cell state, and $\alpha_{B}$ refers to the attentional parameter.
Since the output of the convolution network is input to the LSTM, and the output of the LSTM is the feature map, $x_{t} = x_{j}^{l}$ and $x_{t+1} = x_{j+1}^{l}$.

\subsection{Capsule Networks with Dynamic Routing}
Since the capsule network can effectively learn some aspect features of textual representation~\cite{xiao2018mcapsnet}, the output of the two layers of attentional recurrent convolution networks is $N\times (T-4)\times k2$ size of feature map and is input to the next capsule networks with dynamic routing layer.
In order to independently learn the features of each subgraph into the corresponding capsule vectors, different from existing textual capsule networks~\cite{Zhao2018Investigating}, our proposed capsule networks guarantee the independence of feature between sub-graphs, as shown in Figure~\ref{fig:deepcrcnn}.

The capsules contain groups of locally invariant neurons that learn to recognize the presence of features and encode their properties into vector outputs, with the vector length representing the presence of the features.
The primary capsule layer is a convolution capsule layer with $M$ channels of capsules, as shown in Figure~\ref{fig:deepcrcnn}.
Each primary capsule contains $m$ convolution units with a $\frac{T+12}{9}\times k2$ size of kernel and a vertical stride of 1, and can be seen as the output of all $N\times\frac{T+12}{9}\times k2$ convolution units. 
Here, we guarantee the independence of the representation of sub-graph $\mathcal{G}(v)$.
In total, the primary capsules have $N\times M$ capsule outputs, and each output is a $m$-dimensional vector, as shown in Figure~\ref{fig:deepcrcnn}.
We can see all the primary capsules as a convolution layer with Eq.~\ref{eq:capoutput} as its block non-linearity.
\begin{equation}\label{eq:capoutput}
\begin{aligned}
\centering
v_{j} = \frac{\Arrowvert s_{j} \Arrowvert^{2}}{1 + \Arrowvert s_{j} \Arrowvert^{2}} \cdot \frac{s_{j}}{\Arrowvert s_{j} \Arrowvert},
\end{aligned}
\end{equation}
where $v_{j}$ is the output of capsule $j$, and $s_{j}$ is its total input.
For all but the first layer of capsules, the total input to a capsule $j$ is a weighted sum over all the prediction vectors $\hat{u}_{j|i}$ from the capsules in the layer below, and is calculated by multiplying the output $u_{i}$ of a capsule in the layer below by a weight matrix $W_{ij}$ as following
\begin{equation}\label{eq:weightedsum}
\begin{aligned}
\centering
s_{j} = \sum_{i}c_{ij}\hat{u}_{j|i},\qquad \hat{u}_{j|i} = W_{ij}u_{i},
\end{aligned}
\end{equation}
where $c_{ij}$ is the coupling coefficient that is determined by the iterative dynamic routing process.
The coupling coefficients between capsule $i$ and all the other capsules in the layer above sum to 1. 
They are determined by a \emph{routing softmax} whose initial logits $b_{ij}$ are the log prior probabilities that capsule $i$ should be coupled to capsule $j$.
The $c_{ij}$ can be calculated as following
\begin{equation}
\label{eq:capsoft}
\centering
c_{ij} = \frac{\exp(b_{ik})}{\sum_{k}\exp(b_{ik})}.
\end{equation}
The final DigitCaps layer has $n$ capsules per digit class and each of these capsules receives input from all the other capsules in the layer below. 
$W_{ij}$ is a weight matrix between each $u_{i},i\in (1, M\times N)$ in primary capsules and $v_{j}$, $j\in (1,L)$, where $L$ refers to the number of classes.

As the length of the capsule's output vector represents the presence of a class, the length $\Arrowvert v_{k} \Arrowvert$ of each capsule in the final layer can then be viewed as the probability of the text belonging to a particular class $k$.
The length of the activity vector of each capsule in DigitCaps layer is used to calculate the classification loss.
This encourages the network to learn a more general representation of text with classification task.
Different from the capsule networks~\cite{Sabour2017Dynamic,Hinton2018Matrix,jimenez2018capsule} applied in the field of computational vision, we consider the distance between the raw text and the output of the reconstructed representation in the word embedding space to be relatively large in practice.
We do not perform text reconstruction during training.
Next, we introduce how to design a weighted margin loss to measure distance of classes in hierarchy and guide the training of Attentional Capsule Recurrent CNN model.

\section{Hierarchical Taxonomy-aware Weighted Margin Loss}\label{sec:labelembedding}
Intuitively, the distances between any two classes on the hierarchy are different, but popular margin loss in capsule network~\cite{Sabour2017Dynamic} and other distance measures in multi-label learning~\cite{zhang2014review} between classes didn't consider the hierarchical relations among labels.
So, we explore a hierarchical taxonomy-aware weighted margin loss to guide the training of the proposed model in Section~\ref{sec:dgcrcnn}.

For more formally, we denote the hierarchical taxonomy structure of the labels as $\mathcal{HG} = \{\mathcal{V}, \mathcal{E}\}$, where vertices $\mathcal{V}$ are classes $\mathcal{S}$ and the directed edges $\mathcal{E}$ represent the hierarchical parent-child relationship among the labels. 
In large-scale multi-label text classification, for a document $d_s$ and the corresponding positive labels set $T_s$, the number of labels in $T_s$ is usually much smaller than the remaining negative ones in $S$.
Therefore, it will lead to a large loss of the objective function.
In fact, in hierarchical label network, the closer the edge relationship between nodes, the closer the semantic distance between labels.
In order to conveniently capture the relationship between labels on the hierarchical label structure, we design two meta-paths to guide random walk on the label structure, and generate label sequences to learn label representation.

\begin{figure}[t]
\centering
\includegraphics[width=0.5\textwidth]{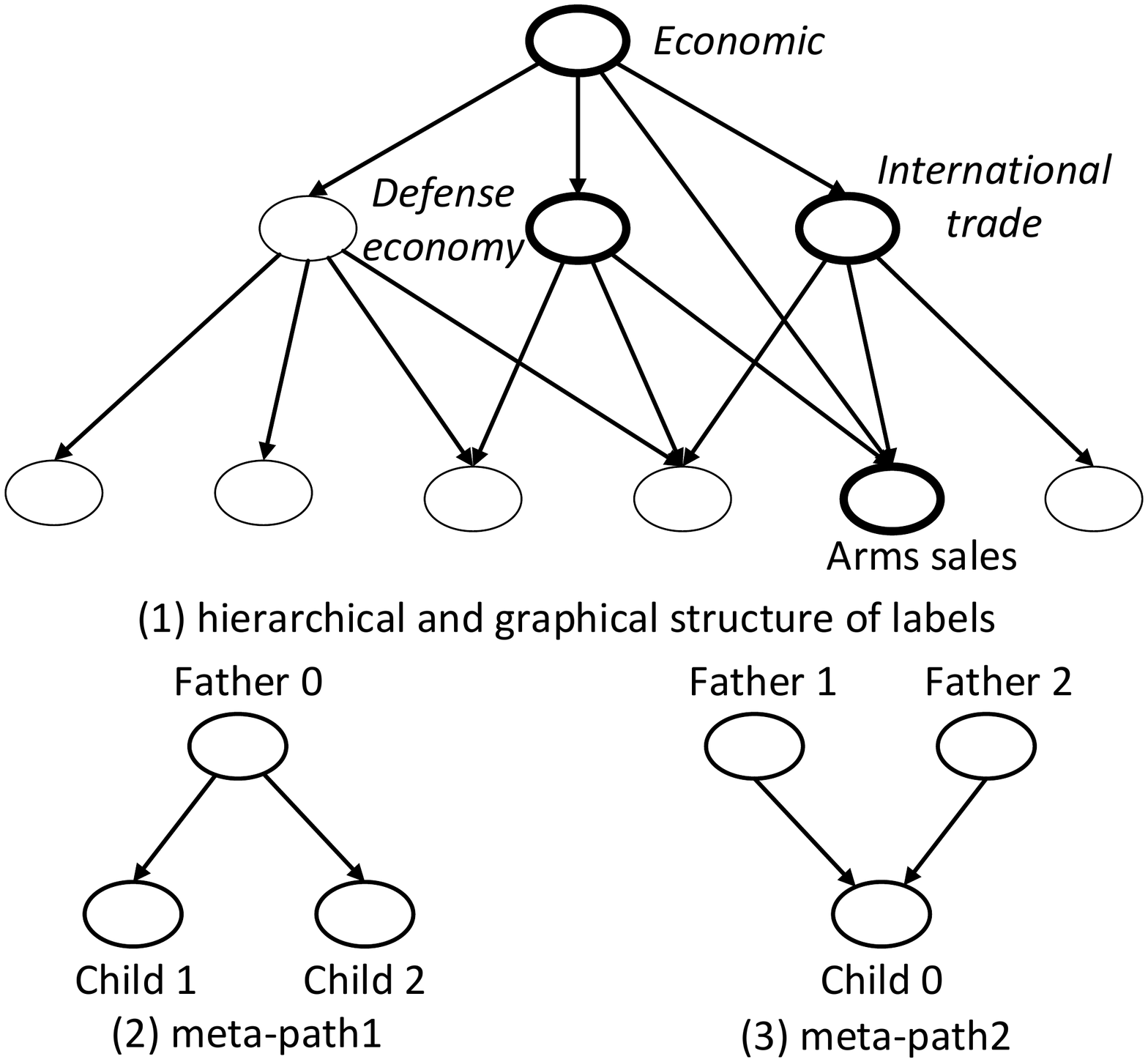}\vspace{-0.3in}
\caption{Illustrations of the hierarchical taxonomy of the labels and the two meta-paths. Vertices represent classes, and edges represent hierarchical parent-child relations.}\label{fig:label2vec}\vspace{-0.2in}
\end{figure}

Figure~\ref{fig:label2vec} illustrates the hierarchical taxonomy structure of the labels, where each node refers to a label/class, and each directed edge represents a parent-child relationship.
Note that the taxonomy network is not a strict hierarchical structure, and may contain cycles. 
For example, the two hierarchical relations of ``\emph{Economic}''-``\emph{International trade}''-``\emph{Arms sales}" and ``\emph{Economic}''-``\emph{Defense economy}''-``\emph{Arms sales}'' can form a cycle.
Both hierarchical structure of taxonomy and graph structure of taxonomy are common in practice, and can be modeled as a hierarchical graph-of-labels. 
For conveniently calculating the distance between any two labels, we measure the discrete cosine distance through their representation vectors. 
As shown in Figure~\ref{fig:label2vec}, we extract the following two types of meta-paths from the hierarchical graph-of-labels, ``\emph{Child1 - Father0 - Child2}'' and ``\emph{Father1 - Child0 - Father2}''.
Actually, the two types of meta-paths control the directions of the random walk to the upper and lower layers, respectively.
So, we perform meta-paths guided random walk to generate sequences of labels.
Here, we set that the probability of selecting two meta-paths is equal during random walking.
Similar to metapath2vec~\cite{dong2017metapath2vec} and Deepwalk~\cite{perozzi2014deepwalk}, we also use the skip-gram with negative sampling ~\cite{Mikolov:2013:DRW:2999792.2999959,Mikolov2013Efficient} to encode the relations among the labels/classes into a continuous vector space.
We optimize the following objective function, which maximizes the log-probability of observing a network neighborhood $N_{S}(l)$ for a node $l$ conditioned on its feature representation, given by $g:$
\begin{equation}\label{eq:node2vec}
\begin{aligned}
\centering
\max_{g} \sum_{l\in V}[-\log Z_{l} + \sum_{n_{i}\in N_{S}(l)}f(n_{i})\cdot f(l)],
\end{aligned}
\end{equation}
where $g(x)=\frac{1}{1+\exp(-x)}$ is the sigmoid function. 
Since it will be time consuming to compute $Z_{l} = \sum_{nl\in V}\exp(g(l)\cdot g(nl))$ for large network, we approximate it using the negative sampling technology.
We optimize the Eq.~\ref{eq:node2vec} by using stochastic gradient method.

Thus, given a tag label/class $l\in \mathcal{V}$, we can approximate a semantic distance between any other label $l_i, i\in [1,L]$ by calculating their discrete cosine distance between their embedding vectors as following:
\begin{equation}\label{eq:disvec}
\begin{aligned}
\centering
d(l,l_i) = 1 - \cos(vec(l),vec(l_i)).
\end{aligned}
\end{equation}
Next, in order to take advantage of dependencies among labels to guide the training of the proposed attentional capsule recurrent CNN model, we design a hierarchical taxonomy-aware weighted margin loss objective function:
\begin{equation}\label{eq:weightedloss}
\begin{aligned}
\centering
\mathcal{L} = \sum_{k=1}^{L} [& T_{k} \max(0, m^{+}-\Arrowvert v_{k}\Arrowvert)^2 \\ & + \lambda \cdot p \cdot \alpha_{k} \cdot (1-T_{k})\cdot \max(0, \Arrowvert v_{k}\Arrowvert - m^{-})^2],
\end{aligned}
\end{equation}
where $T_{k} = 1$ if and only if a digit of class $k$ is present, and $m^{+}$ and $m^{-}$ are the given thresholds for the upper and lower bounds.
The $\lambda$ down-weighting of the loss for absent digit classes stops the initial learning from shrinking the lengths of the activity vectors of all the digit capsules, such as 0.5 in the original capsule networks ~\cite{Sabour2017Dynamic,Hinton2018Matrix}. 
$\alpha_{k} \in [0,1]$ is the minimum semantic distance from negative label $k$ to the positive labels set in the hierarchical label network.
The total loss is the sum of the losses of all the digit capsules.
Formally, for a document $d_{s}$, the positive label set is $T_{s} \subset \mathcal{S}$.
And for any negative label $k$, the $\alpha_{k}$ is:
\begin{equation}\label{eq:params}
\begin{aligned}
\centering
\alpha_{k} = 1 - \underset{t\in T_{s}}{\max}(\cos(vec(t),vec(k))).
\end{aligned}
\end{equation}
Meanwhile, to make an unbiased and smooth overall objective function after integrating the weight, we add an adjustment factor $p$ that satisfies $\sum_{k=1}^{L} p \cdot \alpha_{k} = 1$.
We can approximate the distribution of the semantic distance $\alpha_{k}$ by $1-e^{-x}, x\in [1, L]$ to obtain an approximation of the adjustment factor $p$ for different datasets.

\section{EXPERIMENTS}\label{sec:experi}
In this section, we conduct experiments to evaluate the performance of the proposed framework. 
We will first introduce the used datasets, the evaluation metrics, methods for comparison, and experimental settings. 
Then, we will compare our methods with baselines, and provide the analysis and discussions on the results.

\begin{table*}[h]\caption{\label{tab:data_desc}Statistics of the three datasets. Training, Development, Testing, and Class-Labels denote the total number of training, testing samples and labels, respectively. Words/Sample is the average number of words per sample. Labels/Sample is the average number of labels per sample, and Sample/Labels is the average number of documents per label.}\vspace{-0.15in}
\centering
\begin{tabular}{ccccccccc}
\toprule
Datasets& Training & Development & Testing & Class-Labels& Depth& Words/Sample & Labels/Sample& Samples/Label\\
\midrule
RCV1 & 23,149 & - & 784,446 & 103 & 6 & 268.95 & 3.24 & 729.67\\
EUR-Lex & 15,449 & - & 3,865 & 3,956 & 4 & 1229.77 & 5.32 & 15.59\\
Reuters-21578 & 5800 & 600 & 300 & 10 & - & 257.32 & - & - \\
\bottomrule
\end{tabular}\vspace{-0.1in}
\end{table*}

\subsection{Datasets}
We use two datasets RCV1 and EUR-Lex for large-scale multi-label text classification, and use the Reuters-21578 to evaluate the effectiveness of our proposed capsule network in transferring from single-label to multi-label classification.
The statistics of the datasets is shown in Table ~\ref{tab:data_desc}.

$\bullet$ \textbf{Reuters Corpus Volume I (RCV1)}~\cite{Lewis2004RCV1} is a manually labeled newswire collection of Reuters News from 1996-1997.
It consists of over 800,000 manually categorized newswire stories by Reuters Ltd for research purposes.
Multiple topics can be assigned to each newswire story and there are 103 topics in total.
The news documents are categorized with respect to three controlled vocabularies: industries, topics and regions.
The relations among the labels are typically graphic structure with self-loops. 
We use the topic-based hierarchical classification because it has been widely adopted in evaluation.

$\bullet$ \textbf{EUR-Lex} ~\cite{EUR-LEX} is a collection of documents about European Union law.
It contains many different types of documents, including treaties, legislation, case-law and legislative proposals, which are indexed according to several orthogonal categorization schemes to allow for multiple search facilities. 
The most important categorization is provided by the EUROVOC descriptors, which forms a topic hierarchy with almost 4000 categories regarding different aspects of European law. 
Directory code classes are organized in a hierarchy of 4 levels with a typical tree structure. 
Since the dataset contains several European languages, we choose English version of documents.

$\bullet$ \textbf{Reuters-21578}~\cite{Lewis:1992:EPC:133160.133172,Zhao2018Investigating} is a collection appeared on the Reuters newswire in 1987.
We follow~\cite{Zhao2018Investigating}~\footnote{https://github.com/andyweizhao/capsule\_text\_classification} to choose 6,700 documents from the Reuters financial newswire service, where each document contains either multiple labels or a single label.
And we also focus 10 popular topics, and reprocess the corpus to evaluate the capability of capsule networks of transferring from single-label to multi-label text classification. 
For development and training, we only use the single-label documents in the development and training sets. For testing, we only uses the multi-label documents in testing dataset.
Note that this dataset is only for testing the advantages of the transferring from single-label to multi-label classification task of our capsule network that incorporates multiple semantics.

\subsection{Evaluation Metrics and Baselines}
We use the standard evaluation metrics~\cite{yang1999evaluation} to measure the performance of all the methods.

$\bullet$  \textbf{Micro-$F_{1}$} is a metric considering the overall precision and recall of all the labels.
Let $TP_{t}$, $FP_{t}$, $FN_{t}$ denote the true-positives, false-positives and false-negatives for the $t$-$th$ label in label set $\mathcal{S}$ respectively.
The $Micro$-$F_1$ is defined as:
\begin{align}
\centering
Micro\text{-}F_{1} &=  \frac{2PR}{P+R},
\label{eq:mi-f1}
\end{align}
where:
\begin{align}
& Precision(P) = \frac{\sum_{t\in \mathcal{S}}TP_{t}}{\sum_{t\in \mathcal{S}}TP_{t}+FP_{t}}, \nonumber \\
& Recall(R) = \frac{\sum_{t\in \mathcal{S}}TP_{t}}{\sum_{t\in \mathcal{S}}TP_{t}+FN_{t}}. \nonumber
\end{align}

$\bullet$  \textbf{Macro-$F_{1}$} is a metric which evaluates the averaged $F_1$ of all the different class-labels.
Different from Micro-F1 that gives equal weight to all the instances, $Macro$-$F_{1}$ gives equal weight to each label in the averaging process. 
Formally, $Macro$-$F_1$ is defined as:
\begin{equation}\label{eq:ma-f1}
\begin{aligned}
\centering
Macro-F_{1} =& \frac{1}{|\mathcal{S}|}\sum_{t\in \mathcal{S}}\frac{2P_{t}R_{t}}{P_{t}+R_{t}}, ~~ \text{where} \\
P_{t} =\frac{TP_{t}}{TP_{t}+FP_{t}},& \qquad
R_{t} = \frac{TP_{t}}{TP_{t}+FN_{t}},\\
\end{aligned}
\end{equation}

\begin{table*}[t]\caption{\label{tab:function}Comparison of main functions among variations of HE-AGCRCNN.}\vspace{-0.15in}
	\centering
	\begin{tabular}{c|ccccccc}
		\toprule
	     Models & CNNs & Sorting & LSTM & Attentional LSTM & Capsule & Weighted Margin Loss& \\
		\midrule 
		 TGCNN(No-R) & \checkmark &  \\
         TGCNN & \checkmark & \checkmark\\
         TGRCNN & \checkmark & \checkmark & \checkmark  \\
         GCCNN & \checkmark & \checkmark &  &  & \checkmark \\
         TAGRCNN & \checkmark & \checkmark &  & \checkmark    \\
         GCRCNN & \checkmark & \checkmark & \checkmark &  & \checkmark \\         
         AGCRCNN & \checkmark & \checkmark &  & \checkmark & \checkmark &  \\
         HE-TGCNN & \checkmark & \checkmark & & & & \checkmark \\
         HE-TGRCNN & \checkmark & \checkmark & \checkmark & & & \checkmark \\
         HE-GCCNN & \checkmark & \checkmark & & & \checkmark & \checkmark \\
         HE-TAGRCNN & \checkmark & \checkmark & \checkmark & & & \checkmark  \\
         HE-GCRCNN & \checkmark & \checkmark & \checkmark &  & \checkmark & \checkmark  \\
         HE-AGCRCNN & \checkmark & \checkmark &  & \checkmark & \checkmark & \checkmark  \\
		\bottomrule
	\end{tabular}\vspace{-0.1in}
\end{table*}

Meanwhile, we compare our model with both traditional text classification methods and recent state-of-the-art deep learning based methods.

$\bullet$ \textbf{Flat baselines.} This type of methods generally first extract the TF-IDF features from the document, and then input them into the classification model such as Logistic Regression ({\bf LR}) and Support Vector Machines ({\bf SVM}). 
We call them flat baselines since they ignore both the relations among the words and the relations among the labels, and simply train a multi-class classifier.

$\bullet$ \textbf{N-gram, sequence-of-words or graph-of-words based models.} These methods extract N-gram features, sequence-of-words or graph-of-words from the document as the input of classification models. 
These features are suitable for deep learning models, such as \textbf{CNN-non-static} ~\cite{Kim2014Convolutional}, \textbf{RCNN} ~\cite{Lai:2015:RCN:2886521.2886636}, \textbf{Deep CNN} ~\cite{Conneau2016Very}, \textbf{XML-CNN} ~\cite{Liu:2017:DLE:3077136.3080834}, \textbf{DGCNN-3}~\cite{Peng:2018}, Hierarchical LSTM (\textbf{HLSTM}) ~\cite{chen2016neural}, Hierarchical Attention Network~\cite{Yang2017Hierarchical} (\textbf{HAN}) and Bi-directional Block Self-Attention Network~\cite{shen2018biblosan} (\textbf{Bi-BloSAN}) etc. 
For example, \textbf{HLSTM} model learns sentence representations based on words sequences, and then use RNN models to encode document representations based on the learned sentence representations.
\textbf{HAN} uses a global attention mechanism to attend useful words and sentences.
\textbf{Bi-BloSAN} further splits the sequence into several blocks and employs intra-block and inter-block self-attentions to capture both local and long-range context dependencies, respectively.

$\bullet$ \textbf{Hierarchical models.} These methods make use of the hierarchical or graphical label network to design hierarchical classification classifiers, such as Top-down Support Vector Machines (\textbf{TD-SVM}) ~\cite{Liu:2005:SVM:1089815.1089821}, \textbf{Hierarchical SVMs} ~\cite{Tsochantaridis2006Large}, Hierarchically Regularized Logistic Regression (\textbf{HR-LR}), Hierarchically Regularized Support Vector Machines (\textbf{HR-SVM}) ~\cite{Gopal2013Recursive,Gopal:2015:HBI:2737800.2629585}, and Hierarchically Regularized Deep Graph CNN (\textbf{HR-DGCNN-3})~\cite{Peng:2018}, etc.

$\bullet$ \textbf{Sequence generation model.} These methods view the multi-label classification task as a sequence generation problem, and apply a sequence generation model, such as \textbf{SGM+GE}~\cite{yang2018sgm}, with decoder structure to solve it.

$\bullet$ \textbf{Capsule Neural Networks.} These methods are shown to be effective in capturing the spatial features of text, including Capsule Networks with Dynamic Routing (\textbf{Capsule-A} and \textbf{Capsule-B}) ~\cite{Zhao2018Investigating}. 
These capsule networks rely on N-gram convolution networks to extract shallow features and then use dynamic or static routing to learn the relationships between features. 
\textbf{Capsule-B} model employs parallel networks with different sizes of filters.
It has been proven to have a better effect than the Capsule-A.

$\bullet$ \textbf{Variations of HE-AGCRCNN.} We implement the following several variants of our proposed method. 
Three layers of Graph CNN (\textbf{TGCNN(No-R)}): no-sorting normalization process and without the hierarchical weighted margin loss, attentional LSTM units and capsule network;
Three layers of Graph CNN (\textbf{TGCNN}): without the hierarchical weighted margin loss, attentional LSTM units and capsule network;
Three layers of Graph Recurrent CNN (\textbf{TGRCNN}): without the hierarchical weighted margin loss, attention units and capsule network;
Graph Capsule CNN (\textbf{GCCNN}): without the hierarchical weighted margin loss and attention LSTM units;
Three layers of Attentional Graph Recurrent CNN (\textbf{TAGRCNN}): without the hierarchical weighted margin loss and capsule network;
Graph Capsule Recurrent CNN (\textbf{GCRCNN}): without the hierarchical weighted margin loss and attention units;
Three layers of Attentional Graph Capsule Recurrent CNN (\textbf{AGCRCNN}: without the hierarchical weighted margin loss), and the hierarchical weighted margin loss based models ({\textbf{HE-TGCNN, HE-TGRCNN, HE-GCCNN, HE-TAGRCNN and HE-GCRCNN}}).
All these models have 3 layers of convolutional layers.
In order to clearly present the advantages of the variations of HE-AGCRCNN, we give a table~\ref{tab:function} of models that enhance functionality.

For all the baselines, we use the implementations or open source codes of these models released by authors and other researchers, and report the best performance of the results in our experiments.

\begin{figure*}[t]
\centering
\subfigure[RCV1]
{
    \label{fig:rcv1_embedding}
    \begin{minipage}{0.98\columnwidth}
        \centering
       \includegraphics[width=1\textwidth]{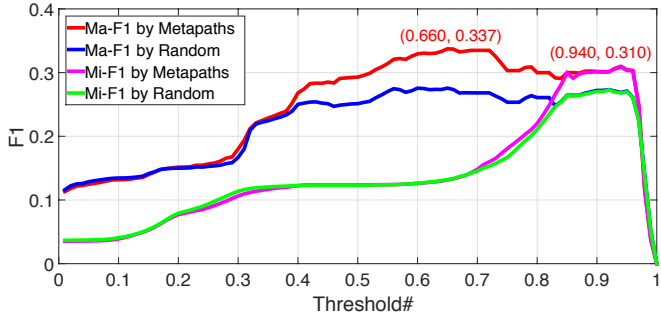}
    \end{minipage}
}\vspace{-0.05in}
\subfigure[EUR-Lex]
{
    \label{fig:EUR_embedding}
   \begin{minipage}{0.98\columnwidth}
       \centering
        \includegraphics[width=1\textwidth]{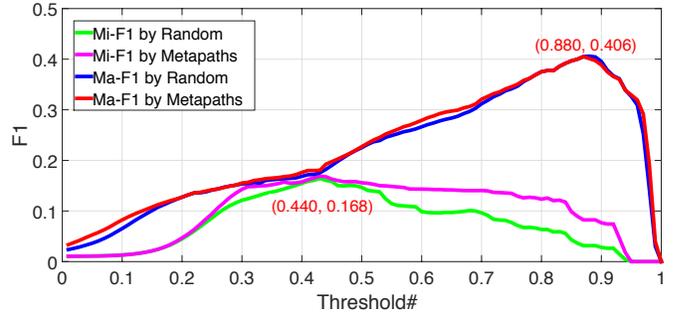}
    \end{minipage}
}\vspace{-0.05in}
\caption{The performance comparisons of reconstructing label network by the two hierarchical taxonomy embedding methods on various thresholds.}\vspace{-0.2in}
\label{fig:embedding}
\end{figure*}

\subsection{Experimental Settings}\label{sec:exp-settings}
All our experiments were performed on 64 core Intel Xeon CPU E5-2680 v4@2.40GHz with 512GB RAM and $8\times$NVIDIA Tesla P100-PICE GPUs.
The operating system and software platforms are Ubuntu 5.4.0, Python 3.5.2, and Pytorch 0.4.0.
The training and testing datasets are shown in Table ~\ref{tab:data_desc}.
In document modeling, the top $N$ numbers of central words are set to 100 (RCV1 and Reuters-21578) and 200 (EUR-Lex).
For the sub-graph, the upper bound value $K$ (the maximum number of vertices) is set to 25.
The length $T$ of the normalized word sequence is set to 20.
The dimension $D$ of word embedding is set to 50.
Here, we use word2vec technology to train 50 dimensional word embedding over the 100 billion words from Wikipedia corpus based on Skip-gram with Negative Sample model with window size of 5.
For the hierarchical taxonomy embedding, we employ 50 threads to execute the random walk in parallel, and for each walk we use 500 steps.
The dimension of label embedding vector is set to 200.
For all the deep learning based models, the common parameters of training the models are empirically set, such as batch size $ = 32$ and learning rate $ = 0.001$ with Adam optimization algorithm.

For the non-capsule neural network models, such as TGCNN(No-R), TGCNN, TGRCNN, TAGRCNN, HE-TGCNN, HE-TGRCNN and HE-TAGRCNN, we use a 2-layer fully connected networks and a sigmoid layer, and the popular cross entropy or the hierarchical taxonomy embedding based weighted margin loss as objective function.
For the capsule network models, we use the the original margin loss or hierarchical taxonomy embedding based weighted margin loss as the objective function.
For capsule based models, the dimension of capsule vector $m$ is 16, the channel of convolution capsule $M$ is 64, the dimensions of DigitCaps are $32\times103$ for the RCV1 dataset, $64\times3956$ for the EUR-Lex, and $32\times10$ for the reprocessed Reuters-21578, and $m^{+} = 0.9$, $m^{-} = 0.1$, $\lambda =0.5$.
Considering the number of the class labels and the average number of labels per sample, we set the adjustment factor $p$ to $0.01$ for RCV1, $0.001$ for EUR-Lex and $0.1$ for the reprocessed Reuters-21578 in the Eq.~\ref{eq:weightedloss}.
All convolution kernels are $1\times3$ in size.
The numbers of convolution kernels per layer are 64 and 128.
The LSTM operator contains 128 hidden layer units.
The numbers of neurons in the fully connected layers are 1024 and 512 in RCV1, and 2048 and 4096 in EUR-Lex.
Our models can achieve the best performance results among 20 to 70 epoches.
For the experiment of transferring single-label model to multi-label classification, on the one hand, in order to be consistent with baseline ~\cite{Zhao2018Investigating}, we select the same number of training, testing, and validation data as shown in Table ~\ref{tab:data_desc}.
On the other hand, as the labels of the reprocessed Reuters-21578 is part of RCV1's, we reuse the vector representation of the RCV1 label of our the hierarchical taxonomy-aware weighted margin loss in the transferring experiment.

\subsection{Evaluation on Label Embedding}
In order to study whether the proposed meta-paths based random walk can learn desirable label embedding that reflects the hierarchical taxonomy relations among them, we use the meta-paths guided random walk and traditional random walk to generate the two label sequences respectively, and then generate two label vectors by the same skip-gram method.
After obtaining the two vectors, we calculate the cosine distance between them, and use it to reconstruct the relations among the labels.
When the distance between two label vectors is larger than the threshold, we add a edge between the two labels. 
We employ the Macro-$F_{1}$ and Micro-$F_{1}$ to evaluate the performance of reconstructing the relations in hierarchy.

Figure~\ref{fig:rcv1_embedding} and figure~\ref{fig:EUR_embedding} show the results of the two label embedding vectors on the relation reconstruction task in the two datasets.
One can see that overall the meta-paths guided random walk approach performs better for capturing the hierarchical taxonomy semantics than the traditional random walk approach.
For RCV1, the most suitable thresholds of meta-paths based taxonomy embedding are $0.660$ and $0.940$, and the highest Macro-$F_{1}$ and Micro-$F_{1}$ are $0.337$ and $0.310$, respectively. 
For the traditional random walk based taxonomy embedding, the highest Macro-$F_{1}$ and Micro-$F_{1}$ are $0.275$ and $0.272$, respectively. 
For EUR-Lex, the most suitable thresholds of meta-paths based taxonomy embedding are $0.440$ and $0.880$, and the highest Macro-$F_{1}$ and Micro-$F_{1}$ are $0.168$ and $0.406$, respectively.
For the traditional random walk based taxonomy embedding, the highest Macro-$F_{1}$ and Micro-$F_{1}$ are $0.164$ and $0.405$, respectively. 
One can observe that the performance difference between the two random walk methods is relatively small on EUR-Lex. 
This is probably because the taxonomy label structure of EUR-Lex is a hierarchical tree.
Although the method of measuring the discrete cosine distance between any two labels based on unsupervised heterogeneous network representation learning vector is approximate, it's a convenient method to estimate label distance for any two labels.

\subsection{Performance Evaluation on RCV1}
Next, we evaluate the performance on the RCV1 dataset through the multi-label text classification task.
RCV1 is a dataset that training samples are much fewer than testing samples, as shown in Table~\ref{tab:data_desc}.

The experiment results are shown in Table~\ref{tab:mltc_results}. 
Among the traditional text classification algorithms, one can see that the HR-SVM performs better than TD-SVM, HSVM, SVM, HR-LR and LR. 
For deep learning approaches, one can see that the performance of RNN based algorithms HLSTM and HAN are comparable to SVM and LR. 
RCNN performs worse on both settings. 
For fine-grained topical classification, the above recurrent models may not have advantages because it compresses the whole document as a dense vector for classification. 
The RNN models are more suitable to sentiment classification for short text, but is not suitable to learn features for long document~\cite{Peng:2018}.
For CNN models, it is shown that XML-CNN does not perform very well on RCV1.
However, the deeper model DCNN improves the performance by 9\% in terms of Macro-F1 and 4\% in terms of Micro-F1.
For capsule network, one can see that the Capsule-B achieves comparable performance with DCNN model.
For sequence generation model, the SGM+GE improves the performance by 2\% in terms of Macro-F1 and Micro-F1 compared with the XML-CNN model.
For GCNN models, both DGCNN-3 and HR-DGCNN-3 improve the performance by 4\% in terms of Macro-F1 and 3\% in terms of Micro-F1 compared with Capsule-B. 
It demonstrates that graph-of-words representation is effective in modeling documents in multi-label text classification.
For the popular bi-directional block self-attention network, the Bi-BloSAN improves the performance by 8\% in terms of Macro-F1 and 3\% in terms of Micro-F1 compared with the HAN model.

For the proposed models, we try different model configurations listed in Table ~\ref{tab:function}.
The results are shown in Table~\ref{tab:mltc_results}.
One can see that the LSTM units, attentional LSTM units, capsule networks and hierarchical taxonomy-aware weighted margin loss are all helpful to improve the classification performance.
The proposed HE-AGCRCNN model outperforms the HR-DGCNN-3 by 8\% in terms of Macro-F1.
The simplified model TGCNN(No-R) also achieves comparable Macro-F1 and Micro-F1 with DGCNN-3.
Meanwhile, without attentional LSTM units, capsule network and hierarchical label dependencies, the TGCNN also outperforms most of the baselines.
Based on the arranged words-matrix representation, LSTM units and attentional LSTM units, the TGRCNN and TAGRCNN models achieve 5\%-6\% improvements in terms of Macro-$F_{1}$ over HR-DGCNN-3.
This improvements show the importance of local sequential semantics for text features. 
Among the proposed models, one can see that capsule networks averagely achieve 1\% gain in both Macro-$F_{1}$ and Micro-$F_{1}$. 
Overall, the hierarchical taxonomy-aware weighted margin loss can also improve the performances by 2\% in terms of Macro-$F_{1}$ and 1\% in terms of Micro-$F_{1}$. 
Finally, the proposed HE-AGCRCNN model achieves the highest $0.513$ Macro-$F_{1}$ and $0.778$ Micro-$F_{1}$ performance.
The results of the different document modeling methods show that by representing document as arranged words-matrix, the proposed model can gain performance improvement for multi-label text classification in RCV1.
One can also see that HR-SVM, HR-DGCNN-3 and HE-AGCRCNN represent two different ways of using the hierarchical label dependencies, and both improve the classification performance over RCV1 dataset.
We will present the timeliness analysis of the experiment in section~\ref{sec:time-analysis}.

\begin{table}[t]\caption{\label{tab:mltc_results}Comparison of results on RCV1 and EUR-Lex.}\vspace{-0.15in}
\centering
\begin{tabular}{c|cc|cc}
\toprule
\multicolumn{1}{c|}{ \multirow{2}*{Models} }& \multicolumn{2}{c|}{RCV1} &\multicolumn{2}{c}{EUR-Lex}\\
\cline{2-5}
\multicolumn{1}{c|}{}& Marco-$F_{1}$& Micro-$F_{1}$&Marco-$F_{1}$& Micro-$F_{1}$\\
\midrule
LR & 0.328 & 0.692 & 0.181 & 0.522\\
SVM & 0.330 & 0.691 & 0.185 & 0.551\\
HSVM &  0.333 & 0.693 & 0.189 & 0.567\\
TD-SVM & 0.337 & 0.696 & 0.198 & 0.571\\
HR-LR & 0.322 & 0.716 & 0.180 & 0.583 \\
HR-SVM & 0.386 & 0.728 & 0.223 & 0.609\\
\hline
HLSTM & 0.310 & 0.673 & 0.183 & 0.562\\
HAN & 0.327 & 0.696 & 0.184 & 0.566\\
RCNN & 0.293 & 0.686 & 0.168 & 0.554\\
XML-CNN & 0.301 & 0.695 & 0.179 & 0.583\\
DCNN & 0.399 & 0.732 & 0.231  & 0.611\\
DGCNN-3 & 0.432 & 0.761 & 0.237  & 0.632\\
HR-DGCNN-3 & 0.433 & 0.762 & 0.241 &  0.649\\
SGM+GE & 0.348 & 0.719 & 0.216 & 0.628\\
Bi-BloSAN & 0.401 & 0.720 & 0.219 & 0.619\\
Capsule-B & 0.399  & 0.739 & 0.226 & 0.600 \\
\hline
TGCNN(No-R) & 0.443  & 0.745 & 0.244 & 0.648\\
TGCNN & 0.472 & 0.747 & 0.257 & 0.655\\
GCCNN & 0.480 & 0.749 & 0.261 & 0.658\\
TGRCNN & 0.484 & 0.754 & 0.265 & 0.667\\
TAGRCNN & 0.490 & 0.759 & 0.270 & 0.673\\
GCRCNN  & 0.488 & 0.765 & 0.275 & 0.668\\
AGCRCNN & 0.494 & 0.769 & 0.283 & 0.675 \\
HE-TGCNN & 0.482  & 0.751 & 0.283  & 0.683\\
HE-GCCNN & 0.491 & 0.754 & 0.290 & 0.688\\
HE-TGRCNN & 0.495 & 0.762 & 0.292 & 0.680 \\
HE-TAGRCNN & 0.504 & 0.773 & 0.298 & 0.685 \\
HE-GCRCNN & 0.505 &  0.772 & 0.297 & 0.684 \\
HE-AGCRCNN & \bf{0.513} & \bf{0.778} & \bf{0.330} & \bf{0.688}\\
\bottomrule
\end{tabular}\vspace{-0.1in}
\end{table}

\subsection{Performance Evaluation on EUR-Lex}
As the number of labels in EUR-Lex is large, we use more neurons in the fully connected layers and set a larger dimension of capsule vector in the DigitCaps layer, as presented in Section~\ref{sec:exp-settings}. 
For the proposed models, we also try different configurations, and the results are shown in Table~\ref{tab:mltc_results}.

From the results one can see that LSTM units, attentional LSTM units, capsule networks and hierarchical taxonomy-aware weighted margin loss are all helpful to improve classification performance on the EUR-Lex dataset. 
HE-AGCRCNN model achieves about 6\% improvements in terms of Macro-$F_{1}$ and 4\% gains in terms of Micro-$F_{1}$ over the HR-DGCNN-3 model.
Without using hierarchical label dependencies, LSTM units, attentional LSTM units and capsule networks, the TGCNN also performs better than HR-DGCNN-3, and the results are $0.257$ and $0.655$ for Macro-$F_{1}$ and Micro-$F_{1}$, respectively.
When we do not order the words in the sub-graph, the results of TGCNN(No-R) are $0.244$ and $0.648$ for Macro-$F_{1}$ and Micro-$F_{1}$, respectively. 
The performance gap between TGCNN(No-R) and TGCNN shows the importance of local sequential semantics for text classification with the same three layers of Graph CNN models. 
Based on the arranged words-matrix representation, the LSTM units can help to improve 1\% performance comparing GCCNN and GCRCNN. 
The performance gap between TGRCNN and TAGRCNN shows that the masked attentional units can help to improve about 0.5\% performance. 
The hierarchical taxonomy-aware weighted margin loss also helps to improve about 1\%-3\%  performances in terms of Macro-$F_{1}$ or Micro-$F_{1}$.
Compared with the improvements of the hierarchical taxonomy-aware weighted margin loss in the RCV1 dataset, the improvements in the EUR-Lex dataset are greater by using the same weighted margin loss.
Finally, HE-AGCRCNN model achieves $0.330$ Macro-$F_{1}$ and $0.688$ Micro-$F_{1}$, which are both the highest performance.
The experimental results again demonstrate that by representing document as an arranged words-matrix and incorporating the proposed deep models, one can gain benefits from non-consecutive, long-distance and sequential semantics for topical multi-label text classification.

\subsection{Performance Evaluation on Reuters-21578}
A significant advantage of capsule network is that it performs much better in the transferring single-label to multi-label classification task~\cite{Zhao2018Investigating}.
Different from traditional deep learning classification models that are based on fully connected network, capsule networks use activity vectors of each capsule in DigitCaps layer to indicate the presence of an instance of each class.
We also perform the model transfer capacity experiment of the proposed capsule network on the reprocessed Reuters-21578 dataset~\cite{Zhao2018Investigating}.

The comparison results are shown in Table ~\ref{tab:RCV-transferring}. 
The baseline results are also reported from the work~\cite{Zhao2018Investigating}, and HE-AGCRCNN outperforms all the baselines.
Compared with capsule-based models, the performances of LSTM, BiLSTM and CNN-non-static are the worst.
From the results one can see that GCCNN, GCRCNN, AGCRCNN, HE-GCCNN, HE-GCRCNN and HE-AGCRCNN models all have achieved about 5\%-7\% improvements in terms of Micro-$F_{1}$ over the existing best baseline Capsule-B.
Even without the attentional LSTM units, the simplified GCCNN can achieve $0.905$ performance in terms of Micro-$F_{1}$.
Compared with the popular Capsule-A and Capsule-B models, our proposed GCCNN, GCRCNN, AGCRCNN, HE-GCCNN, HE-GCRCNN and HE-AGCRCNN models integrate more non-consecutive, long-distance and local sequential semantics, and make use of the hierarchical label dependencies.
Finally, the proposed HE-AGCRCNN model achieves the highest $0.927$ performance in terms of Micro-$F_{1}$.
The experimental improvements again prove the effectiveness of our proposed capsule models in learning rich textual features.

\begin{table}[t]\caption{\label{tab:RCV-transferring}Comparison of the transferring capacity from single-label to multi-label text classification on the reprocessed Reuters-21578 dataset.}\vspace{-0.15in}
\centering
\begin{tabular}{c|ccc}
\toprule
Models& P & R & Micro-$F_{1}$\\
\midrule
LSTM & 0.867 & 0.547 & 0.635\\
BiLSTM & 0.823 & 0.559 & 0.643\\
CNN-non-static & 0.920  &  0.597 &  0.704\\
Capsule-A & 0.882 & 0.801 & 0.820\\
Capsule-B & 0.954 & 0.820 & 0.858\\
\hline
GCCNN & 0.962 & 0.856 & 0.905 \\
GCRCNN & 0.970 & 0.871 & 0.917 \\
AGCRCNN & 0.973 & 0.875  & 0.921 \\
HE-GCCNN & 0.965 & 0.862 & 0.910 \\
HE-GCRCNN &  0.974 & 0.879 & 0.924 \\
HE-AGCRCNN & \bf{0.978} & \bf{0.882} & \bf{0.927}\\
\bottomrule
\end{tabular}\vspace{-0.1in}
\end{table}

\subsection{Training Efficiency Evaluation}\label{sec:time-analysis}
To evaluate the training efficiency of the model, we next show the training time of the proposed model and its variants on both RCV1 and EUR-LEX datasets in Table~\ref{tab:GPU}. 
Here, for the RCV1, since the number of samples in the test set is about 34 times larger than the number of samples in the training set, as shown in Table~\ref{tab:data_desc}, we perform the testing by using multi-core CPUs.

One can observe that most of these models can quickly achieve a promising classification result with less than 3 hours except for the models with the LSTM or capsule unites. 
For example, the TGCNN and HE-TGCNN models converge quickly with less than 0.2 hour on RCV1 dataset and less than 1.2 hours on EUR-Lex dataset. 
Meanwhile, the training time on RCV1 dataset is much less than EUR-Lex. 
This is mainly because the EUR-Lex dataset has a larger document representation and more parameters, according to Table ~\ref{tab:data_desc}. 
We also verify that the models integrating more feature extraction operators, such as LSTM units, attentional LSTM units and capsule networks, will take longer time to train for achieving a desirable classification performance.
Although HE-AGCRCNN model takes $1.119$ hours and $6.335$ hours to train for RCV1 and EUR-Lex datasets, respectively, it achieves the highest classification performance. 
One can also see that the hierarchical taxonomy embedding based weighted margin loss does not add much computational time compared with the recursive regularized optimization models~\cite{Peng:2018,Gopal2013Recursive,Gopal:2015:HBI:2737800.2629585}.
Usually, the time consumptions of the above recursive regularization based models are expensive for the large number of parameters and constraints on the Euclidean distance of the parameters.
In particular, the time consumptions of recursive regularization optimized deep learning model, such as HR-DGCNN-3, is generally measured in days~\cite{Peng:2018}.

\begin{table}[t]\caption{\label{tab:GPU}Comparison of training time on GPUs.}\vspace{-0.15in}
	\centering
	\begin{tabular}{c|cc}
		\toprule
	     Models& RCV1(hr.) & EUR-Lex(hr.)\\
		\midrule 
		 TGCNN & 0.166 & 1.100 \\
         GCCNN & 0.537 & 3.415 \\
         TGRCNN & 0.381 & 2.579 \\
         TAGRCNN & 0.382 & 2.580 \\
         GCRCNN & 1.116 & 6.327 \\
         AGCRCNN & 1.117 & 6.328  \\
         HE-TGCNN & 0.167 & 1.167 \\
         HE-GCCNN & 0.542 & 3.421 \\
         HE-TGRCNN & 0.385 & 2.583 \\
         HE-TAGRCNN & 0.386 & 2.584  \\
         HE-GCRCNN & 1.118 & 6.334 \\
         HE-AGCRCNN & 1.119 & 6.335 \\
		\bottomrule
	\end{tabular}\vspace{-0.1in}
\end{table}

\begin{figure*}[t]
\center
\includegraphics[width=1\textwidth]{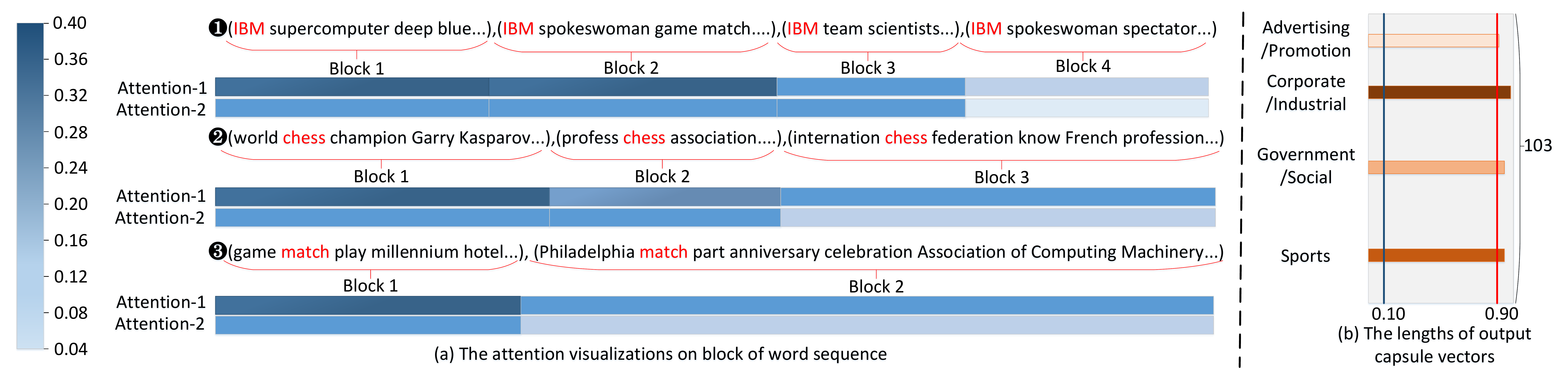}\vspace{-0.3in}
\caption{The attention visualizations and output capsule vectors for the 3093newsML sample in RCV1. The left one is parts of attention visualizations on blocks of word sequences, and the right one is the length of output capsule vectors. }\label{fig:casestudy}\vspace{-0.2in}
\end{figure*}

\subsection{Case study}
To gain a closer view of what's the two attention layers and output capsules in a document captured by our models, we visualize parts of the attention probability or alignment score by heatmaps in Figure~\ref{fig:casestudy}.
The red words are central words in the document, each block of word sequence is context of central words.
For each central word, there are two layers of masked attention.
We choose the blocks of word sequences from the $1$-th, $2$-th and $3$-th central words.
One can see that the weights of the upper left parts are higher than other places.
This is probably because the contextual semantics of the front central words are more representative of the subject of the article.
For the output capsule vectors, there are 4 vectors whose modulus length is greater than 0.9, corresponding to the category of output.
Note that Sports category comes out due to the words game, match, defeat and win, although chess is not really a physical activity sport.

\section{Related Work}\label{sec:relatedwork}
As our work is closely related to text classification, textual deep learning models and graph convolution networks, in this section we will review related works from the three aspects.

Tradition text classification models use feature engineering and feature selection to obtain features for text classification~\cite{Aggarwal2012}.
For example, Latent Dirichlet Allocation~\cite{BleiNJ03} has been widely used to extract “topics” from corpus, and then represent documents in the topic space.
It performs better than Bag-Of-Word (BOW) when the feature numbers are small.
However, when the size of words in vocabulary increases, it does not show advantage over BOW on text classification ~\cite{BleiNJ03}.
There are also some existing work that tried to convert texts to graphs ~\cite{Rousseau2015Text}.
Similar to our proposed methods, they used word co-occurrence to construct graphs from texts, and then they applied similarity measure on graph to define new document similarity and features for text ~\cite{Rousseau2015Text}.
For hierarchical large-scale multi-label text classification, many efforts have been put on how to leverage the hierarchy of labels to improve the classification results.
Recently, a recursive regularization of weight euclidean constraint with classifiers has been developed, and shown to be the out-performance in large-scale hierarchical text classification problems~\cite{Gopal2013Recursive,Gopal:2015:HBI:2737800.2629585}.

For deep learning models, there have been RNNs, CNNs, and capsule models applied to text classification.
For example, hierarchical RNN has been proposed for long document classification~\cite{TangQL15} and later attention model is also introduced to emphasize important sentences and words~\cite{Yang2017Hierarchical}.
Similar to RNNs, the recently proposed self-attention based sentence embedding technologies~\cite{lin2017structured,shen2018disan,shen2018biblosan} have shown effectively capturing both long-range and local dependencies in sentiment-level tasks.
For example, Bi-BloSAN~\cite{shen2018biblosan} is a bi-directional block self-attention network to learn text representation and models text as sequences.
For CNNs models, Kalchbrenner et al. ~\cite{kalchbrenner2014convolutional} and Kim et al.~\cite{Kim2014Convolutional} used simpler CNN for text classification, and showed significant improvements over traditional texts classification methods.
Zhang et al. ~\cite{Zhang2015Character} and Conneau et al. ~\cite{Conneau2016Very} used a character level CNN with very deep architecture to compete with traditional BOW or n-gram models.
The combination of CNNs and RNNs are also developed which shows improvements over topical and sentiment classification problems ~\cite{Lai:2015:RCN:2886521.2886636}.
Capsule networks were proposed by Hinton et.~\cite{Hinton2011Transforming,Sabour2017Dynamic,Hinton2018Matrix} as a kind of supervised representation learning methods, in which groups of neurons are called capsules.
Capsule network has been proved effective in learning the intrinsic spatial relationship between features~\cite{Zhao2018Investigating,xiao2018mcapsnet,zhang2018attention}.
~\cite{Zhao2018Investigating} showed that Capsule networks can help to improve low-data and label transfer learning.
However, as mentioned in the introduction, existing textual deep learning models are not compatible with diverse text semantic coherently learning.
Compared with our work, these previous studies only considered N-gram or sequential text modeling, but ignored high level of non-consecutive and long-distance semantics of text. 
They did not study and utilize the dependency among the the labels, either.
Although there are prior works~\cite{Gopal2013Recursive,Peng:2018,xie2013multilabel,garg2015exploring} on modeling pair-wise relation between labels for multi-label classification, they fail to consider their hierarchical relations, and the computation of the above models is expensive due to the use of euclidean constraints in their regularization.

GCN derived from graph signal processing ~\cite{Shuman2013The,Bruna2013Spectral}, and the graph convolution operation has been recognized as the problem of learning filter parameters that were replaced by a self-loop graph adjacency matrix, updating network weights, and extended by utilizing fast localized spectral filers and efficient pooling operations in  ~\cite{Defferrard2016Convolutional,Kipf2016Semi,Duvenaud2015Convolutional}.
With the development of GCN technologies, graphs embedding approaches, such as PSCN~\cite{Niepert:2016:LCN:3045390.3045603} and GCAPS-CNN~\cite{Verma2018Graph}, have been developed in graph classification tasks.
Recently, the recursively regularized deep graph-cnn~\cite{Peng:2018} has been proposed to combine graph-of-words representation, graph CNN, and hierarchical label dependency for large-scale text classification.
Then the Text GCN model~\cite{YaoGCN2018} has been proposed to capture global word co-occurrence information and perform text classification without word embeddings or other external knowledge.
Although long-distance and non-continuous text features are fully considered in the two models, the existing graph convolutional neural network models ignore the continuous and sequential semantics of words in the text.
In addition, the recursive regularization is usually time consuming due to the euclidean constraint.

\section{Conclusion and Future Work}\label{sec:conclu}
In this paper, we present a novel end-to-end hierarchical taxonomy-aware and attentional graph capsule recurrent CNN framework for large-scale multi-label text classification. 
We first propose to convert each document as an arranged words-matrix that preserves both the non-consecutive, long-distance and local sequential semantics for fully representing the document. 
Based on our document modeling, we next propose a HE-AGCRCNN model to coherently learn multiple types of textual features.
In order to better learn local sequential semantics, we design a masked attentional LSTM to model the different impacts among different blocks of word sequences, and enhance the sequential features learning.
To incorporate the hierarchical relations among the labels, we further propose a novel hierarchical taxonomy-aware weighted margin loss to improve the performance of multi-label text classification.
The advantageous performance of our proposed models over other competing methods is evident as it obtained the best results on all the RCV1 and EUR-Lex datasets in our comparative evaluation.
Compared to the N-gram based textual capsule networks, we verify the effectiveness of our proposed capsule models in learning rich textual features in transferring single-label to multi-label classification task.
The experimental results show the effectiveness and efficiency of our model in multi-label text classification.

In the future, we plan to invest subgraph-level attention capsule network, and upgrade our hierarchical taxonomy-aware and attentional graph capsule recurrent CNN to self attention rnn/cnn models~\cite{shen2018biblosan,shen2018disan}, variable-size convolution kernels~\cite{yin2015multichannel} and BERT~\cite{devlin2018bert} pre-trains based models, and popularize to more sophisticated text classification datasets and applications.

\bibliography{newref}

\begin{IEEEbiography}
[{\includegraphics[width=1in,height=1.2in,clip,keepaspectratio]{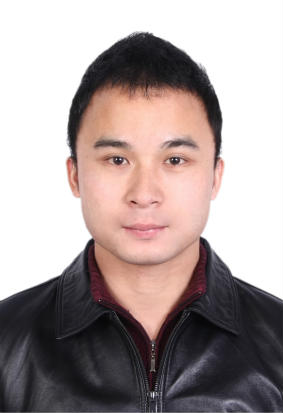}}]
{Hao Peng} is currently a Ph.D. candidate at the State Key Laboratory of Software Development Environment, and Beijing Advanced Innovation Center for Big Data and Brain Computing in Beihang University. His research interests include representation learning, text mining and urban computing.
\end{IEEEbiography}
\vspace{-0.3in}

\begin{IEEEbiography}
[{\includegraphics[width=1in,height=1.2in,clip,keepaspectratio]{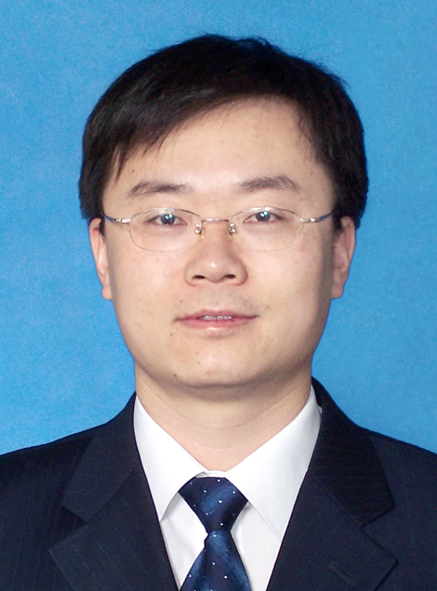}}]
{Jianxin Li} is currently a Professor with the State Key Laboratory of Software Development Environment, and Beijing Advanced Innovation Center for Big Data and Brain Computing in Beihang University. His current research interests include social network, machine learning, distributed system, virtualization, big data, trust management and network security.
\end{IEEEbiography}
\vspace{-0.3in}

\begin{IEEEbiography}
[{\includegraphics[width=1in,height=1.2in,clip,keepaspectratio]{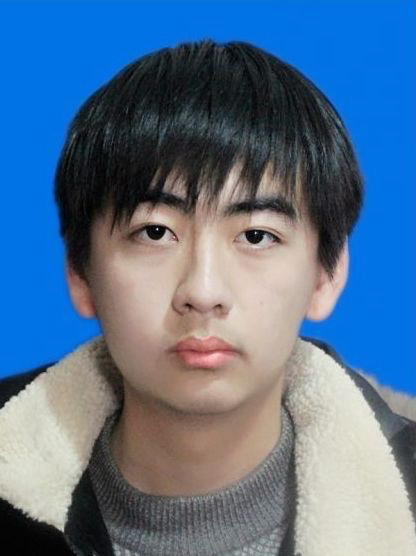}}]
{Qiran Gong} is currently a B.E. candidate at the State Key Laboratory of Software Development Environment in Beihang University, Beijing, China. His research interests include social network mining and text mining.
\end{IEEEbiography}
\vspace{0.2in}

\begin{IEEEbiography}
[{\includegraphics[width=1in,height=1.2in,clip,keepaspectratio]{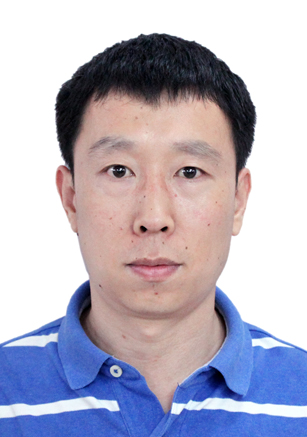}}]
{Senzhang Wang} is currently an Associate Professor with the Collage of Computer Science and Technology, Nanjing University of Aeronautics and Astronautics, Nanjing. His current research interests include data mining, urban computing and social network analysis.
\end{IEEEbiography}
\vspace{-0.3in}

\begin{IEEEbiography}
[{\includegraphics[width=1in,height=1.2in,clip,keepaspectratio]{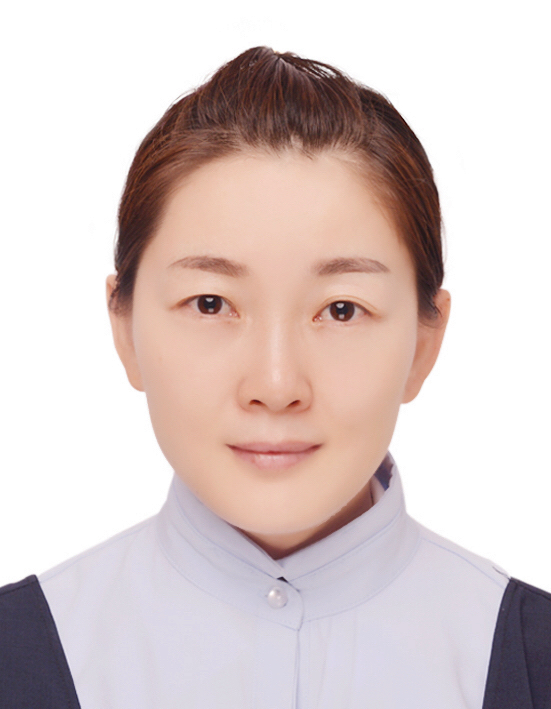}}]
{Lifang He} is currently a Postdoctoral Research Associate at the Department of Biostatistics and Epidemiology at the University of Pennsylvania. Her current research interests include machine learning, data mining, tensor analysis, biomedical informatics.\par
\end{IEEEbiography}
\vspace{-0.3in}

\begin{IEEEbiography}
[{\includegraphics[width=1in,height=1.2in,clip,keepaspectratio]{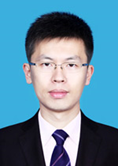}}]
{Bo Li}, is currently an Associate Professor with the State Key Laboratory of Software Development Environment, and Beijing Advanced Innovation Center for Big Data and Brain Computing in Beihang University. His current research interests include big data computing theory, machine learning and computer security.
\end{IEEEbiography}
\vspace{-0.3in}

\begin{IEEEbiography}
[{\includegraphics[width=1in,height=1.2in,clip,keepaspectratio]{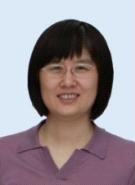}}]
{Lihong Wang} is a professor in National Computer Network Emergency Response Technical Team/Coordination Center of China. Her current research interests include information security, cloud computing, big data mining and analytics, information retrieval and data mining.
\end{IEEEbiography}
\vspace{-0.3in}

\begin{IEEEbiography}
[{\includegraphics[width=1in,height=1.2in,clip,keepaspectratio]{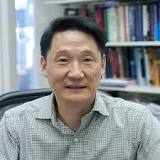}}]
{Philip S. Yu} is a Distinguished Professor and the Wexler Chair in Information Technology at the Department of Computer Science, University of Illinois at Chicago. Before joining UIC, he was at the IBM Watson Research Center, where he built a world-renowned data mining and database department. He is a Fellow of the ACM and IEEE. Dr. Yu is the recipient of ACM SIGKDD 2016 Innovation Award for his influential research and scientific contributions on mining, fusion and anonymization of big data, the IEEE Computer Society’s 2013 Technical Achievement Award for “pioneering and fundamentally innovative contributions to the scalable indexing, querying, searching, mining and anonymization of big data” and the Research Contributions Award from IEEE Intl. Conference on Data Mining (ICDM) in 2003 for his pioneering contributions to the field of data mining. Dr. Yu has published more than 1,100 referred conference and journal papers cited more than 103,000 times with an H-index of 152. He has applied for more than 300 patents. Dr. Yu was the Editor-in-Chiefs of ACM Transactions on Knowledge Discovery from Data (2011-2017) and IEEE Transactions on Knowledge and Data Engineering (2001-2004).
\end{IEEEbiography}

\end{document}